\documentclass{kluwer}    

\newdisplay{guess}{Conjecture}

\begin{document}                                                                                   
\begin{article}
\begin{opening}         
\title{Holographic View of Cosmological Perturbations } 
\runningtitle{Holographic View of Cosmological Perturbations}
\author{Jiro \surname{Soda}}
\runningauthor{Jiro Soda \& Sugumi Kanno}
\institute{Department of Fundamental Sciences, FIHS, 
Kyoto University, Kyoto 606-8501, Japan}
\author{Sugumi \surname{Kanno}}  
\institute{ Graduate School of Human and Environmental
 Studies, Kyoto University, Kyoto
 606-8501, Japan }


\begin{abstract}
 The cosmological perturbation theory is revisited from the holographic
  point of view. 
 In the case of the single brane model, it turns out that the AdS/CFT 
 correspondence plays an important role. 
  In the case of  the two-brane model, it is shown that 
  the effective equations of motion  becomes  
  the quasi-scalar-tensor gravity. It is also demonstrated that 
  the radion anisotropy gives the CMB fluctuations through the Sachs-Wolfe 
  effect.   
\end{abstract}


\end{opening}           

\section{Introduction}  

 In this paper, we would like to propose a new approach
  to the cosmology
 in the context of the warped compactification
 modeled by the action~\cite{RS1,RS2}
\begin{equation}
S = {1\over 2\kappa^2}\int d^5 x \sqrt{-g} \left( 
       {\cal R} + {12\over l^2} \right)  
       -\sigma \int d^4 x \sqrt{-h} \\
  + \int d^4 x \sqrt{-h} {\cal L}_{\rm matter} \ ,
\end{equation}
where $\sigma,l, \kappa^2 $ are the brane tension,  AdS  curvature scale, 
 and the gravitational constant, respectively. Here,   
$ h_{\mu\nu}$ denotes the induced metric on the brane

In the brane world cosmology, in general, we must consider the  initial 
and  boundary conditions at the same time~\cite{koyama1,koyama2}. 
For this reason, it is difficult to deduce 
the predictions such as CMB fluctuations. 
 To circumvent this situation, we solve the 5 dimensional bulk dynamics firstly,
  and obtain the  4-dimensional effective theory. To carry out this 
 program, we must adopt the low energy approximation.
 The procedure we performed will be referred to as holographic projection 
 in this paper~\cite{KS1,KS2}. 
 The point is that the  low energy approximation is useful 
because  $\rho \leq (1{\rm TeV})^4 ({1{\rm cm} / l})^2$ is not so low!

\section{Holographic Projection}

 Let us start with a review of the geometrical approach~\cite{SMS}.  
In the  Gaussian normal coordinate system: 
$
 ds^2 = dy^2 + g_{\mu\nu} (y,x^{\mu} ) dx^{\mu} dx^{\nu}   \ ,
$
we have  
\begin{eqnarray}
 G^{(5)}_{\mu\nu}  &=& G^{(4)}_{\mu\nu} + K_{\mu\nu,y} - g_{\mu\nu} K_{,y}
   - K K_{\mu\nu} + 2 K_{\mu\lambda} K^{\lambda}_{\nu} \nonumber \\
  & & \qquad +{1\over 2} g_{\mu\nu} 
  \left( K^2 + K^\alpha_\beta K^\beta_\alpha \right)
   = { 6\over l^2} g_{\mu\nu}     \ ,
\end{eqnarray}
where   $K_{\mu\nu} = -  g_{\mu\nu ,y} /2 $.
 Taking into account the $Z_2$ symmetry,   we also obtain the junction 
 condition 
$
 \left[ K^\mu_\nu - \delta^\mu_\nu K \right] |_{y=0} 
    = {\kappa^2 / 2}  \left( -\sigma \delta^\mu_\nu + T^\mu_\nu \right) 
$. 
Here, $T_{\mu\nu}$ represents the 
energy-momentum tensor of the matter. 
Evaluating Eq.(2) at the brane and substituting the junction condition 
into it, we have the ``effective" equations of motion 
\begin{eqnarray}
 G^{(4)}_{\mu\nu} &=& 8\pi G_N T_{\mu\nu} + \kappa^4 \pi_{\mu\nu} 
                   - E_{\mu\nu} \\
  && \pi_{\mu\nu} = -{1\over 4}T_\mu^\lambda T_{\lambda\nu} 
  +{1\over 12}TT_{\mu\nu} + {1\over 8} g_{\mu\nu} \left( 
   T^{\alpha\beta} T_{\alpha\beta} -{1\over 3} T^2 \right) \nonumber \\
  && E_{\mu\nu} = C_{y\mu y \nu} |_{y=0}  \nonumber \ ,
\end{eqnarray}
where $C_{y\mu y \nu}$ is the Weyl tensor and $8\pi G_N = \kappa^2 /l$. 
We also assumed $\kappa^2 \sigma = 6/ l $. 

 Notice that the above geometrical projection is not the closed system.
  Our aim  is to get the closed system of the equations.
 The above Eq.(3)
  can be transformed into a more convenient form for this purpose. 
  In fact, the equation  
$
 \nabla^{\mu} E_{\mu\nu} = \kappa^4 \nabla^{\mu} \pi_{\mu\nu}
$
 derived from the Bianchi identity can be integrated to
\begin{equation}
  E_{\mu\nu} = \kappa^4 \pi_{\mu\nu} 
  - l^2 \chi_{\mu\nu}  - t_{\mu\nu}
\end{equation}
where $\nabla^\mu \chi_{\mu\nu}=0$ and $\nabla^\mu t_{\mu\nu} =0$.
 Here, we have divided the integration constant
 into the nonlocal part $\chi_{\mu\nu}$ and the local part $t_{\mu\nu}$.
 The explicit formula for $\chi_{\mu\nu}$ and $t_{\mu\nu}$ can 
  be obtained by resorting to the low energy approximation.  
  
  The low energy approximation can be reformulated as the 
  gradient expansion as can be seen from the estimate: 
$
  {\rho / \sigma}   \sim  l^2 R \ll 1  \ .
$
In the gradient expansion, we can expand $\chi_{\mu\nu}$ and $t_{\mu\nu}$
 as
\begin{eqnarray}
  \chi_{\mu\nu} &=& \underbrace{\chi^{(1)}_{\mu\nu}}_{
       \mbox{$ {\cal O} (l^2 R) $}} +\underbrace{\chi^{(2)}_{\mu\nu}}_{
       \mbox{$ {\cal O} (l^4 R^2 ) $}} + \cdots \\
        t_{\mu\nu} &=& \qquad \qquad \underbrace{t^{(2)}_{\mu\nu}}_{
       \mbox{$ {\cal O} (l^4 R^2 ) $}} + \underbrace{t^{(3)}_{\mu\nu}}_{
       \mbox{$ {\cal O} (l^6 R^3 ) $}}  + \cdots    \ ,
\end{eqnarray}
where $t_{\mu\nu}$ can be obtained from $h_{\mu\nu}$ and its derivatives. 
The property $E^\mu_{\ \mu} =0$ leads the following relations
\begin{eqnarray}
      \chi^{(1)\mu}_{\quad \mu} &=& 0 \\
   l^2 \chi^{(2)\mu}_{\quad \mu} &=& \kappa^4 \pi^{\mu}_{\ \mu} 
   - t^{(2)\mu}_{\ \mu}  \ .
\end{eqnarray}

Having obtained $\chi_{\mu\nu}$ and $t_{\mu\nu}$, we 
 finally get the holographic projection  
\begin{equation}
     G^{(4)}_{\mu\nu} = 8\pi G_N T_{\mu\nu} 
     +   l^2 \chi_{\mu\nu} 
         + t_{\mu\nu }  \ .
\end{equation}
 It should be noted that
 the bulk metric can be reconstructed perturbatively
 from the data obtained by solving Eq.(9).
 That is why we use the terminology ``holography"  for 
 our projection method.

\section{Single brane model (RS2): AdS/CFT Correspondence}

The nonlocal tensor $\chi_{\mu\nu}$ must be related to the 
   boundary conditions in the bulk. 
 The natural choice is  asymptotically  AdS boundary condition.
 For this boundary condition,  $\chi^{(1)}_{\mu\nu} =0 $. Hence, 
  Einstein theory is recovered at the leading order!

 At ${\cal O} ( l^4 R^2 ) $ order, one can take $t^{(2)\mu}_{\quad \mu} =0 $.
Indeed, the gradient expansion method gives
\begin{eqnarray}
  \alpha t^{(2)\mu}_{\quad \nu} &=& R^\mu_{\ \alpha} R^\alpha_{\ \nu}
             -{1\over 3} R R^\mu_{\ \nu} 
         -{1\over 4} \delta^\mu_\nu (R^\alpha_{\ \beta} R^\beta_{\ \alpha}
         - {1\over 3} R^2)   \\ 
    & & \qquad     -{1\over 2} \left( R^{\alpha\mu}_{\ \ |\nu\alpha}
                    + R^{\alpha \ |\mu}_{\ \nu \ \  | \alpha}  
              -{2\over 3} R^{|\mu}_{\ |\nu}  - \Box R^\mu_{\ \nu} 
              +{1\over 6} \delta^\mu_\nu \Box R \right)   \nonumber
\end{eqnarray}
which is transverse and traceless, 
$  t^{(2)\mu}_{\quad \nu | \mu} =0  \ ,  t^{(2)\mu}_{\quad \mu} = 0 \ .  
$
Moreover,  using Eq.(8) and the lowest order equation 
$    T^\mu_{\ \nu} 
   \approx {l / \kappa^2 } (R^\mu_{\ \nu} - {1\over 2} \delta^\mu_\nu R) 
   \ ,   $ 
we have 
\begin{equation}
  \chi^{(2)\mu}_{\quad \mu}  
     ={1\over 4} \left( R^\alpha_{\ \beta} R^\beta_{\ \alpha} 
            - {1\over 3}  R^2 \right)     \ .  
\end{equation}
Utilizing the AdS/CFT correspondence, we can identify 
\begin{equation}
\chi^{(2)}_{\mu\nu} = { \kappa^2\over l^3} <T_{\mu\nu}^{\rm CFT}>  \ . 
\end{equation}
Thus, we obtain the holographic effective equations of motion
\begin{equation}
     G^{(4)}_{\mu\nu} = 8\pi G_N T_{\mu\nu} + 8\pi G_N <T_{\mu\nu}^{\rm CFT}>
         + \alpha  t^{(2)\mu}_{\quad \nu }  \ .
\end{equation}

Now we can consider the cosmology. The background spacetime
 is nothing but the ordinary
 one since the correction does not affect the isotropic homogeneous 
 background.  
 It is the cosmological perturbations that we are interested in.
 From the renormalized action for the CFT, $S^{\rm CFT}$, we have
\begin{equation}
     <T^{\rm CFT}_{\mu\nu}> = -<{2\over \sqrt{-g}} 
     {\delta S^{\rm CFT} \over \delta g^{\mu\nu}} >
\end{equation}
and  
\begin{equation}
   <T^{\rm CFT}_{\mu\nu} (x) T^{{\rm CFT}\lambda\rho}(y) > 
       = -{2\over \sqrt{-g}} {\delta <T^{\rm CFT}_{\mu\nu}>
        \over \delta g^{\lambda\rho}}   \ .
\end{equation}
Hence, the perturbed effective Einstein  equations are obtained as
\begin{eqnarray}
  \delta G_{\mu\nu} &=& 8\pi G \delta T_{\mu\nu}  \\
     && \quad  -{1\over 2} \int d^4 y \sqrt{-g(y) } 
       <T^{\rm CFT}_{\mu\nu} (x) T^{{\rm CFT}\lambda\rho}(y) >
        \delta g_{\lambda\rho}
       + \alpha \delta t^{(2)}_{\mu\nu}  \ . \nonumber 
\end{eqnarray}
This is the integro-differential equation. 
 Notice that  $<T^{\rm CFT}_{\mu\nu} (x) T^{{\rm CFT}\lambda\rho}(y) > $ 
are given  kernel, once the  CFT is specified. Moreover,  
 $\alpha$ is a parameter determined by the initial conditions for 
the bulk geometry.
 Therefore, in principle, we can obtain the time evolution of
  the cosmological  perturbations by solving Eq.(16).

\section{Two-brane model (RS1): Radion}

 We consider  the two-brane system in this section.
 At the lowest order of the low energy approximation, 
 we have the background  metric
\begin{equation}
ds^2 = e^{2 \phi(x)} dy^2 + \exp[- {2\over l}  e^{\phi(x) } y] 
                                h_{\mu\nu} (x) dx^\mu dx^\nu  \ ,
\end{equation}
where the radion field $\phi$ is related to the distance between two branes as
$ d(x) =  e^{\phi(x) } l $.

 At the next order, the holographic projection gives
\begin{eqnarray}
     G^{(4)}_{\mu\nu} (h_{\mu\nu} )&=& {\kappa^2 \over l} T^A_{\mu\nu} 
                          + l^2 \chi^{(1)}_{\mu\nu}  \\
     G^{(4)}_{\mu\nu} (f_{\mu\nu})&=& - {\kappa^2 \over l}  T^B_{\mu\nu} 
               + l^2 \chi^{(1)}_{\mu\nu} \exp[4 e^{\phi (x)}] \ ,
\end{eqnarray}
where $f_{\mu\nu}$ is the induced metric on the $B$-brane.
 The extra factor attached to the last term in Eq.(19)
 can be deduced from the 5-dimensional equations. 
 It should be noted that 
$h_{\mu\nu}$ and $f_{\mu\nu}$ are not independent but related as 
$f_{\mu\nu} = \exp[- 2 e^{\phi (x)}] h_{\mu\nu} $ at this order.  
 Therefore, we can eliminate $\chi_{\mu\nu}$ from
 Eqs.(18) and (19), and obtain the quasi-scalar-tensor gravity.
Indeed, defining a new field $\Psi = 1- \exp[ -2 e^{\phi(x)}] $,  we find 
\begin{eqnarray}
 	G^\mu_{\ \nu} (h) &=&{\kappa^2 \over l \Psi } T^{A\mu}_{\quad\ \nu}
      		+{\kappa^2 (1-\Psi )^2 \over l\Psi } T^{B\mu}_{\quad\ \nu}
      		+{ 1 \over \Psi } \left(  \Psi^{|\mu}_{\ |\nu} 
  		-\delta^\mu_\nu  \Psi^{|\alpha}_{\ |\alpha} \right)  
  		                        \nonumber\\
  	& &	+{\omega(\Psi ) \over \Psi^2} \left( \Psi^{|\mu}  \Psi_{|\nu}
  		- {1\over 2} \delta^\mu_\nu  \Psi^{|\alpha} \Psi_{|\alpha} 
  		\right)  \ ,
\end{eqnarray}
\begin{equation}
  	\Box \Psi = {\kappa^2 \over l} {T^A + (1-\Psi) T^B 
  	                                \over 2\omega (\Psi) +3}
  		-{1 \over 2\omega (\Psi) +3}
  		{d\omega (\Psi) \over d\Psi} \Psi^{|\mu} 
  		\Psi_{|\mu}  \ ,
\end{equation}
where the coupling function $\omega (\Psi)$ takes the  form:
$
  	\omega (\Psi ) = 3 \Psi / 2(1-\Psi )   
$.
This action can be a starting point of various applications such as the 
inflation in the two-brane system. 

 Importantly, we can also deduce the formula for Weyl fluid
\begin{eqnarray}
 {l^3 \over 2}  \chi^{(1)\mu}_{\quad \nu} 
 &=& -{\kappa^2 (1-\Psi) \over 2 \Psi} 
      		\left( T^{A\mu}_{\quad \nu} 
      		+ (1-\Psi ) T^{B\mu}_{\quad \nu} \right) \\
      	      & &  -{l  \over 2 \Psi} \left[ \left(  \Psi^{|\mu}_{\ |\nu} 
  		-\delta^\mu_\nu  \Psi^{|\alpha}_{\ |\alpha} \right)
  		+{\omega(\Psi ) \over \Psi} \left( \Psi^{|\mu}  \Psi_{|\nu}
  		- {1\over 2} \delta^\mu_\nu  \Psi^{|\alpha} \Psi_{|\alpha} 
  		\right) \right]    \nonumber \ .
\end{eqnarray}
The explicit form for the Weyl fluid is now known provided
 the effective equations (20) and (21) are solved.

\section{Effect on CMB fluctuations}

 Let us briefly touch on a possible effect of bulk gravitational waves 
 on CMB fluctuations.
 Taking the perturbed 4-dimensional metric  as 
\begin{equation} 
ds^2 = -(1+2\phi) dt^2 + (1+2 \psi) \delta_{ij} dx^i dx^j \ ,
\end{equation}
and  defining the gauge invariant curvature perturbation  
\begin{equation} 
\zeta = \psi + {\delta \rho \over 3 (\rho + p)} \ ,
\end{equation}
we have the Sachs-Wolfe formula for the CMB temperature 
anisotropy\\~\cite{Lang}
\begin{equation}
  {\delta T \over T}  = \zeta + \psi - \phi 
  + \int d\eta {\partial \over \partial \eta} (\psi - \phi ) \ .
\end{equation}
 In order to give a precise value, we must know the anisotropic 
 stress due to the Weyl fluid. 
For two brane system, the Weyl fluid is represented by the radion field
 as in Eq.(22). Hence, the  formula (25) reduces to
\begin{eqnarray}
  {\delta T \over T}|_{\rm SW} &=& {\delta T \over T}|_{4D} 
     - {8\over 3} {\rho_r \over \rho_d } S_{\rm Weyl} \\
    & &  +{1 \over \Psi_0} 
    \left( \delta\Psi^{|i}_{|j} -{1\over 3} \delta \Psi^{|k}_{|k} 
                                      \right)
          - {2 \over a^{5/2}  } \int {da \over \Psi_0 } a^{3/2}
           \left( \delta \Psi^{|i}_{|j} -{1\over 3} \delta \Psi^{|k}_{|k} 
                                      \right) \nonumber\ .
\end{eqnarray}
Here, the radion dynamics is important to predict the observational 
consequences. As we have the effective theory already, 
the evolution of the radion field can be calculated easily.
 Then, the remaining issue is to determine the initial conditions.
 The initial value for $\delta\Psi$ should be determined by 
 calculating quantum fluctuations in the inflationary period. 
 This is now under investigation.

\section{Conclusion}

We revealed the holographic aspects of the brane world cosmology.
 In the case of the single brane model, 
 imposing the asymptotically AdS boundary condition, 
 we obtained the Einstein equations with  CFT and  higher curvature corrections.
 As to the  two-brane model, it is shown that  
Einstein  equations with the  nonlocal Weyl fluid
 are converted  to the quasi-scalar-tensor gravity. 
 There,  the  radion field played an important role. 
 In particular, the Weyl fluid is explicitly written down
 using the radion field. Consequently, we succeeded to write down 
 the  formula for CMB anisotropy
 ${\delta T / T}$ in a closed form.  
 The precise predictions will be published in the future.

\acknowledgements
\theendnotes
We would like to thank M. Sasaki for valuable suggestions. 
This work was supported in part by  Monbukagakusho Grant-in-Aid No.14540258.

\end{article}
\end{document}